\documentclass[twocolumn,prl,showpacs,preprintnumbers,amsmath,amssymb]{revtex4}
\usepackage{graphicx}% Include figure files
\usepackage{dcolumn}% Align table columns on decimal point
\usepackage{bm}% bold math

\newcommand{\ket}[1]{|#1\rangle}
\newcommand{\MgH}[0]{$^{24}\text{MgH}^+$ }
\newcommand{\Mg}[0]{$^{24}\text{Mg}^+$ }
\newcommand{\Trot}[0]{$\text T_{\text {rot}}$ }

\begin{document}

\preprint{APS/123-QED}

\title{The rotational temperature of polar molecular ions in Coulomb crystals}% Force line breaks with \\

\author{Anders Bertelsen}
\affiliation{%
QUANTOP - Danish National Research Foundation Center for Quantum
Optics, Department of Physics and Astronomy, University of Aarhus,
DK-8000 Aarhus C, Denmark
}%

\author{Solvejg J\o rgensen}
\affiliation{%
QUANTOP - Danish National Research Foundation Center for Quantum
Optics, Department of Physics and Astronomy, University of Aarhus,
DK-8000 Aarhus C, Denmark
}%

\author{Michael Drewsen}
\affiliation{%
QUANTOP - Danish National Research Foundation Center for Quantum
Optics, Department of Physics and Astronomy, University of Aarhus,
DK-8000 Aarhus C, Denmark
}%

\date{\today}

\begin{abstract}

With MgH$^+$ ions as a test case, we investigate to what extent
the rotational motion of smaller polar molecular ions
sympathetically cooled into Coulomb crystals in linear Paul traps
couples to the translational motions of the ion ensemble. By
comparing results obtained from rotational resonance-enhanced
multiphoton photo-dissociation experiments with data from
theoretical simulations, we conclude that the effective rotational
temperature exceeds the translational temperature ($<$ 100 mK) by
more than two orders of magnitude, indicating a very weak
coupling.

\end{abstract}

\maketitle

%\section{Introduction}

Molecules which are translationally cold, spatially localized and
internally prepared in specific states constitute ideal targets
for a large variety of investigations, including high resolution
spectroscopy \cite{Hollas}, coherently controlled chemistry
\cite{opticalcontrol}, state-specific reaction studies
\cite{Harich, Liu}
 and molecular internal state
control experiments \cite{stapelfeldt:543}.

For atoms such targets have for years been available by combining
laser cooling techniques with trapping and, perhaps most
spectacularly, been exploited in quantum logics experiments
\cite{Riebe2004,Barrett2004}. It is, however, extremely difficult
to adopt the same techniques for molecules due to the general lack
of closed optical transitions. For polar molecules a further
complication arises since the rovibrational degrees of freedom
couple rather efficient to the black body radiation (BBR) of the
surroundings. Recently, various methods that at least partially
approach the ideal situation have, however, been demonstrated for
neutral molecules. These include photo-association of cold atoms
\cite{fioretti:4402}, buffer gas cooling of molecules in magnetic
traps \cite{Weinstein}, and control of the motion of molecules by
either by static \cite{Bethelm6732606} or optical electrical
fields \cite{fulton:243004}.
%
%and deceleration and trapping of dipolar molecules by electrical
%fields \cite{bethlem:1558}. Most recently, even molecular BEC of
%molecules has been created \cite{Jochim2003}.
%
%
%
%Molecular ions sympathetically cooled \cite{larson:70} through
%Coulomb interactions with laser cooled atomic ions in Penning or
%RF traps is an alternative route to achieve a near-ideal
%experimental situation. Since this method does not rely on an
%electric or a magnetic dipole moment of the molecular ion, it is
%very generally applicable. So far, this technique has proven it
%possible to cool either single or ensembles of molecular ions into
%translational temperatures in the mK-range \cite{drewsen:243201,
%Moelhave, schiller:053406} where they become part of a Coulomb
%crystal.
%

Internal cold, but not spatial localized molecular ions have been
produced either by photo-processes \cite{Muller2395176, zhu:5769},
by buffer-gas cooling in RF multi-pole traps \cite{Gerlich,
brummer:12700} or by state selective recombination
\cite{lammich:143201}. Recently, by sympathetical cooling to
translational temperatures in the mK-range in RF ion traps,
spatially localized molecular ions in Coulomb crystals have been
reported \cite{Moelhave, drewsen:243201}.

In this Letter, we investigate whether there is a sufficient
coupling between the internal degrees of freedom of polar
molecular ions and the external motion of the ions in a Coulomb
crystal to achieve a significant sympathetic cooling of the
internal states of the molecular ions. The strategy is to apply
Resonance-Enhanced Multiphoton Photo-Dissociation (REMPD)
\cite{VanHeijnsbergen7469220} as a tool to obtain knowledge of the
populations in the various rotational states and hence determine
the effective rotational temperature. One could most likely expect
a measurable cooling of the internal degrees of freedom of the
molecular ions due coupling of the dipole moment of the molecules
to the motion of the charges of the ions in the Coulomb crystal.
However, since the typical frequencies of the normal modes of the
Coulomb crystal will be $\sim 10^6$ Hz while the rovibrational
frequencies will be $\sim 10^{11}$ Hz -- $\sim 10^{13}$ Hz, no
strong resonant coupling is expected Ref. \cite{LangArtikkel}.

%\section{How to determine $T_\text{rot}$}

An ensemble of \MgH molecules, which is translational
sympathetically cooled by laser cooled \Mg atomic ions in a linear
Paul trap has been chosen as a test case for many reasons. The
\MgH ions are easy to produce and sympathetically cool
\cite{Moelhave}, at room temperature 99\% will be in the
vibrational ground state, precise spectroscopic data are available
for \MgH \cite{Balfour1972:1082}, and finally in order to compare
experimental results with theoretical simulations, the potential
curves and permanent as well as transition dipole moments of \MgH
can be calculated with good precision
\cite{Solvejg,vogelius:173003}.

%An ensemble of \MgH molecules, which is translational
%sympathetically cooled by laser cooled \Mg atomic ions in a linear
%Paul trap has been chosen as a test case for many reasons. First,
%these molecular ions have previously been found easy to produce
%and sympathetically cool \cite{Moelhave}. Second, the vibrational
%constant of \MgH is sufficiently large that the molecules even at
%room temperature will have a 99\% vibrational ground state
%population, which means that we experimentally can focus on the
%rotational degree of freedom. Third, in order to implement the
%REMPD process it is a great advantage that precise spectroscopic
%data are available for \MgH \cite{Balfour1972:1082}. Fourth, in
%order to compare experimental results with theoretical
%simulations, the \MgH molecule has a sufficiently simple structure
%that potential energy curves and permanent as well as transition
%dipole moments can be calculated with good precision
%\cite{Solvejg,vogelius:173003}. Finally, the \MgH ion is of
%atmospheric and astrophysical interest \cite{SinghSol}.
\begin{figure}
%    \psfrag{C}[][][2]{C$^1\Sigma$}
%    \psfrag{B}[][][2]{B$^1\Pi$}
%    \psfrag{A}[c][c][2]{A$^1\Sigma$}
%    \psfrag{X}[][][2]{X$^1\Sigma$}
%    \psfrag{ytitle}[][][2.3]{Potential energy [eV]}
%    \psfrag{xtitle}[][][2.3]{Intermolecular distance [\AA]}
%    \psfrag{as}[][][2]{\hspace{0.5in}$\ket{ \nu=0,\text J }_\text X$ }
%    \psfrag{cs}[][][2]{\hspace{0.5in}$\ket{ \nu',\text J'}_\text X$ }
%    \psfrag{bs}[][][2]{\hspace{0.5in}$\ket{ \nu''=0,\text J+1 }_\text A$ }
%    \psfrag{af}[c][c][2.5]{$\varepsilon$}
%    %\psfrag{bf}[][][1.7]{\hspace{0.1in}$\varepsilon_b$}
%    \psfrag{disX}[][l][2]{\hspace{0.41in}Mg$^+$(3s)+H(1s)}
%    \psfrag{disA}[][l][2]{\hspace{0.41in}Mg$^+$(3p)+H(1s)}
%    \psfrag{disC}[][l][2]{\hspace{0.41in}\hspace{-0.11in}Mg(3s$^2$)+H$^+$}
%\psfrag{Ge}[][l][2]{$\Gamma_\text E$}
% \psfrag{Gvj}[ ][t][2]{$\Gamma_{\nu \text{J}} $}
    \resizebox{0.85\columnwidth}{!}{\includegraphics{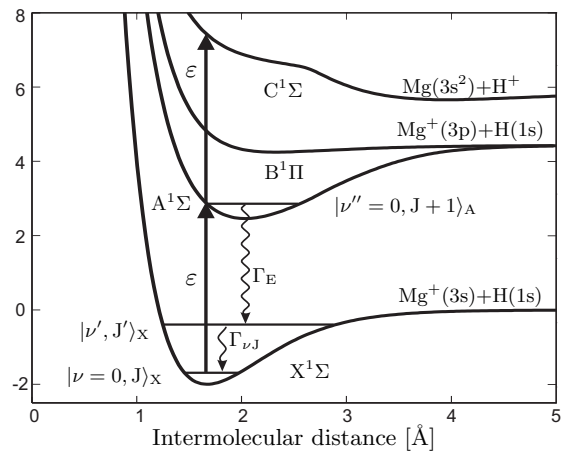}} %schemefig1b_ver2.eps}}%rottempLett_080605pic1.eps}}  %schemefig1b.eps}} rottempLett_120405pic1.eps}} %
 \caption{The four lowest spin singlet potential curves for \MgH including various symbols and processes discussed in the text.}\label{fig. 1}
\end{figure}

In Fig. \ref{fig. 1}, the energy potential curves of the four
lowest lying spin singlet electronic states of \MgH are shown
together with an indication of states and processes of relevance
to the investigations. The resonant part of the two-photon REMPD
process consists of a rotational state-selective excitation from a
state $\ket{ \nu=0,\text J }_\text X$ in the X$^1\Sigma$ potential
to a state $\ket{ \nu''=0,\text J+1 }_\text A$ in the A$^1\Sigma$
potential by a laser field $\varepsilon$. The letters in the
ket-notation represent the vibrational and rotational quantum
numbers, respectively. From the $\ket{ \nu''=0,\text J+1 }_\text
A$ state, the molecule can dissociate nearly exclusively into Mg
and H$^+$ via the C potential curve by absorbing another photon
from the $\varepsilon$ field. However, with the field strengths
applied in our experiments, the excited molecule primarily
spontaneously decays back to a vibrational excited state in the
electronic ground state potential X$^1\Sigma$ with a rate
$\Gamma_\text E\approx 3\times 10^{8}$ $ \text s^{-1}$.
Subsequently, further spontaneous emissions, stimulated processes
due to the coupling to the BBR field and  transitions induced by
an effective coupling to the external motion of the ions take
place. In Fig. 1, these various mechanisms are in brief denoted by
$\Gamma_{\nu \text J}$. Eventually, the molecule will reach the
state $\ket{ \nu=0,\text J }_\text X$ from where it again can
couple strongly to the laser field $\varepsilon$.

%\section{Experimental issues}

\begin{figure}
%    \psfrag{a}[][][2]{\White{(a)}}
%        \psfrag{b}[][][2]{\White{(b)}}
%            \psfrag{c}[][][2]{\White{(c)}}
 \resizebox{0.85\columnwidth}{!}{\includegraphics{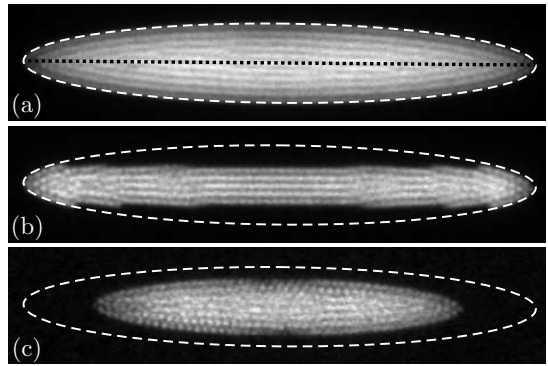}}\\ %krystalpic_17_260704_2.eps}}\\ %krystalpic_17_260704_1.eps}}\\
  \caption{\textbf{(a)} 2D projection-image of a pure Coulomb crystal of approximately
3000 fluorescing \Mg ions. The length of the crystal is 1070 $\mu
$m.
 \textbf{(b)} A two-component crystal, consisting of $\sim35\%$ \Mg and $\sim65\%$ \MgH ions (non-fluorescing), obtained by letting \Mg ions in the crystal in (a) react with a gas of H$_2$.
\textbf{(c)} A nearly pure \Mg crystal obtained after ns-laser
pulses have been applied for 60 s to dissociate the \MgH molecules
in the crystal shown in (b). The dashed ellipse indicates
 the position of the outer surface of the initial crystal in (a). The dotted line in (a) indicates the axis of rotational symmetry.
}\label{fig. crystal pic}
\end{figure}

The experimental setup has previously been described in Refs.
\cite{drewsen:Massspec2003, bertelsen}. In short, \Mg ions are
produced by off-resonance photo-ionization by the pulsed laser
field $\varepsilon$ indicated in Fig. 1 and trapped in a linear
Paul RF trap \cite{drewsen:Massspec2003} situated in a UHV chamber
at background pressure of $\sim 1 \times 10^{-10}$ torr. The
produced \Mg ions are laser-cooled on the
3s$^2$S$_{1/2}$-3p$^2$P$_{3/2}$ transition by two cw laser beams
counter-propagating along the center-axis of the trap to
temperatures below 10 mK where they form a Coulomb crystal
\cite{drewsen:Massspec2003}. The two-dimensional projection images
of such crystals are obtained by monitoring the fluorescence by a
CCD camera. A typical projection image of a pure \Mg crystal is
shown in Fig. \ref{fig. crystal pic}(a). The shape of the real
crystal is spheroidal with rotational symmetry around the dotted
line.

The \MgH molecular ions are formed through photochemical reactions
between \Mg ions excited to the 3p$^2$P$_{3/2}$ state and H$_2$
molecules leaked in the chamber at partial background pressure of
$5\times 10^{-10}$ torr for a few minutes \cite{Moelhave}. The
produced \MgH molecular ions are within tens of milliseconds
translational sympathetically cooled by the remaining \Mg to a
temperature below 100 mK, and a two-component Coulomb crystal is
formed \cite{Moelhave}. As seen in Fig. \ref{fig. crystal pic}(b),
due to dynamical radial confinement of the ions, the
non-fluorescing \MgH molecular ions and the coolant \Mg ions
segregate radially with the lighter mass ions, the \Mg ions,
closest to the rotational symmetry axis
\cite{drewsen:Massspec2003}. The amount of each ion species in the
crystal can be deduced from the fact that each species in the trap
will have a uniform density inversely proportional to its mass and
that for crystals containing two species with a small difference
in mass, the outer boundary will have a near spheroidal shape
\cite{bertelsen, hornekaer:1994, schiller:053406}. By determining
the volume of the fluorescing region in Fig. \ref{fig. crystal
pic}(b), we can conclude that this particular crystal has a
relative \MgH content of $\sim$ 65\%.

The laser field $\varepsilon$ presented in Fig. \ref{fig. 1}
originates from a frequency doubled dye laser pumped by a Nd:YAG
laser. The pulse repetition rate is 20 Hz, and the duration of
each pulse is approximately 5 ns. The waist of the field at the
trap center is 500 $\mu$m, and the peak intensity was typically of
the order of I$_{\text{peak}}\sim 10^7$ W/cm$^2$, corresponding to
a peak power-broadening of the $\ket{\nu=0,\text J}_\text
X$--$\ket{\nu''=0,\text J+1 }_\text A$ transitions of $\sim0.3$
cm$^{-1}$.
The spectral width of each pulse was $\sim$ 0.25 cm$^{-1}$, and
the center frequency was calibrated to better than 0.1 cm$^{-1}$.
During the experiments, the center frequency was scanned 0.95
cm$^{-1}$ at 1 Hz in order to obtain results which are independent
of slow frequency drifts in the laser system. The resonance
frequencies of the X$^1\Sigma(\nu=0,\text J)$--A$^1\Sigma(
\nu''=0,\text J + 1)$ transitions, which are used in our REMPD
scheme, are known with a precision of 0.05 cm$^{-1}$ from Ref.
\cite{Balfour1972:1082}.

%\subsection{Data presentation}
Before the REMPD lasers are applied, the produced \MgH ions are allowed
to internally equilibrate for about one minute, a time long enough
to ensure that the produced molecular ions are in the vibrational
ground state and that any effective cooling of the
rotational degrees of freedom will be revealed in the experiments.
In the experiments, crystals like the one shown in Fig. 2(b) are
exposed to the laser pulses while CCD images are being recorded at
a rate of 6 Hz. After a typical exposure time of tens of seconds,
all molecules have been dissociated, and a nearly pure crystal of
\Mg is left in the trap as shown in Fig. \ref{fig. crystal
pic}(c). Before a new REMPD experiment is carried out, the trap is reloaded
with \Mg ions and through reactions, Coulomb crystals with nearly
the same number of \Mg and \MgH ions are created each time.

\begin{figure}
% \psfrag{Ytitle}[][][2]{Normalized \MgH amount}
%    \psfrag{Xtitle}[][][2]{Exposure time [s]}
   \resizebox{0.85\columnwidth}{!}{\includegraphics{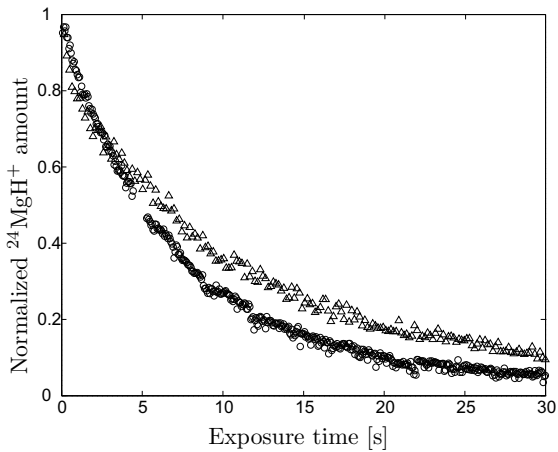}}\\ %R1R6amountWithoutDecayRatesAndFits.eps}}\\
  \caption{The amount of \MgH ions relative to the initial content of \MgH in the crystal as a function
   of the time in which the \MgH ions are exposured to the dissociation laser pulses.
    The triangles and circles represent data for the R(1) and R(6) resonances, respectively (see text). In this case I$_{\text{peak}}=1.9 \times 10^7$ W/cm$^2$.}\label{Fig. 3}
\end{figure}

In Fig. \ref{Fig. 3}, the normalized amount of \MgH molecular ions
are shown for two out of about ten REMPD experiments where the
$\varepsilon$ field has been tuned to the resonance of the
R-branch transitions R(1): $\ket{\nu=0,\text J=1 }_\text
X$--$\ket{\nu''=0,\text J= 2 }_\text A$ and R(6):
$\ket{\nu=0,\text J=6 }_\text X$--$\ket{\nu''=0,\text J= 7 }_\text
A$, respectively.

%%%%%%%%%%%%%solvjg start
In order to derive an estimate for the rotational temperature from
the outcome of the REMPD experiments, simulations including
various strengths of a effective sympathetic rotational cooling
have been performed. In the model, this cooling has been accounted
for by introducing an artificial decay rate from the rotational
states J to J-1 proportional to J. This corresponds roughly to
have an effective cooling rate $\Gamma_{c}$ defined through
$d\overline{E}_{rot}/dt=-\Gamma_{c}\overline{E}_{rot}$, where
$\overline{E}_{rot}$ is the average rotational energy. In Fig. 4,
the simulated steady-state rotational temperature of the \MgH
ions, defined as \Trot$=\overline{E}_{rot}/k_B$, is presented as a
function of the cooling rate $\Gamma_{c}$. In the absence of
sympathetic cooling, the rotational distribution is observed to
equilibrate to the temperature of the BBR field in the trap region
($\sim$ 300 K). It is, however, clear that only a very modest
cooling rate $\Gamma_{c}$ ($\sim 0.1 $ $\text s^{-1}$) will change
this situation dramatically (see Fig. 4). The steady-state
rotational distributions are the stating points for all
simulations.

When simulating the time evolution of the photo-dissociation
process, the time scale is broken into two types of intervals
defined by the presence or absence of the laser pulses. During a
laser pulse, the effect of laser driven transitions between the
various rovibrational states of the X$^1\Sigma$ and A$^1\Sigma$
potential curves is taken into account by at any time during the
pulse assuming that the various populations are in steady-state
with the instantaneous laser field strength and detunings. Based
on the instantaneous populations in the A$^1\Sigma$ potential
curve, the dissociation rate via the C$^1\Sigma$ potential curve
is determined. During the laser pulse, spontaneous emission from
the populated A$^1\Sigma$ states to the various rovibrational
states of the X$^1\Sigma$ potential curve is included as well,
while rovibrational transitions within the potential curves are
neglected due to the short time scale of the pulses. Between
consecutive pulses population redistribution processes among the
rovibrational states in the electronic ground potential
X$^1\Sigma$ are simulated taking into account a effective
sympathetic cooling, spontaneous emissions and interactions with
the BBR field. Due to the complexity of the simulations, only the
main ideas have been considered here, but the details will be
presented elsewhere Ref. \cite{LangArtikkel}.

%%%%%%%%%%%solvejg end
\begin{figure}
%    \psfrag{Ytitle}[][][2]{Rotational temp. T$_{\text{rot}}$ [K]}
%    \psfrag{Xtitle}[][][2]{Cooling rate $\Gamma_c$ [s$^{-1}$]}
%    \psfrag{xtitle}[][][2]{J}
%    \psfrag{ytitle}[][][2]{Population}
   \resizebox{0.89\columnwidth}{!}{\includegraphics{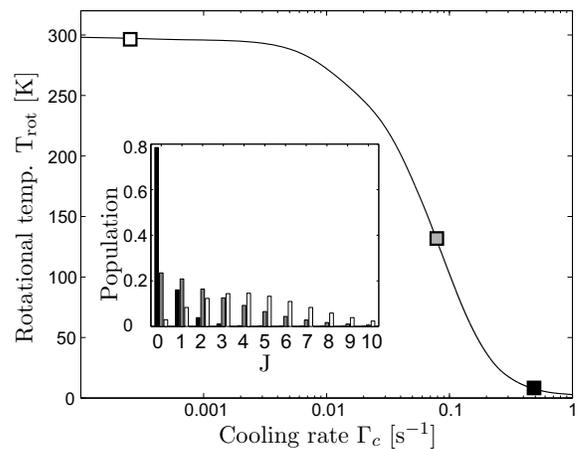}}\\ %TrotvsCoupling_withsplineand_insert.eps}}\\ %cool_T1}}\\
  \caption{The steady-state rotational temperature of \MgH as a function of the
   effective cooling rate $\Gamma_c$ (see text for detailed definitions).
    The insert shows the population distributions for three different rotational
     temperatures: White: \Trot=300 K, Grey: \Trot=120 K, and Black: \Trot=7 K.
      The colors of the bars correlates to the color of the three squares in the main plot.}\label{Fig. 4}
\end{figure}
%\section{Comparison of experiment and simulations}

\begin{figure}
%    \psfrag{Ytitle}[][][2]{Ratio of dissociation rates}
%    \psfrag{Xtitle}[][][2]{Rotational temp. T$_{\text{rot}}$ [K]}
%    \psfrag{a}[c][r][1.7 ]{I$_\text{peak}=$3.80$\times 10^7$W/cm$^2$}
%    \psfrag{b}[c][r][1.7 ]{I$_\text{peak}=$1.90$\times 10^7$W/cm$^2$}
%    \psfrag{c}[c][r][1.7]{I$_\text{peak}=$0.95$\times 10^7$W/cm$^2$}
  \resizebox{0.85\columnwidth}{!}{\includegraphics{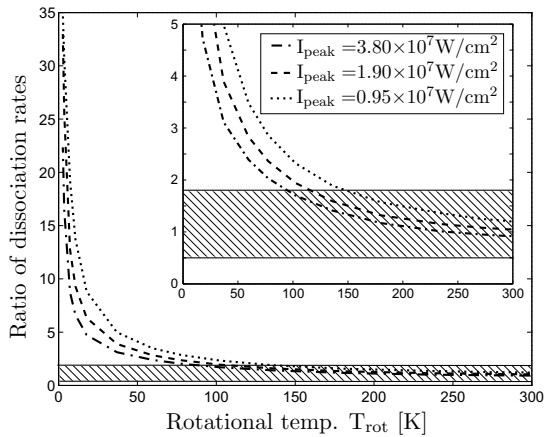}}\\ %simR1R6rateratiowithZoom.eps}}\\ %Sim_R6vsR1_artjan}}\\
  \caption{The ratios between the simulated dissociation rates for REMPD
processes via the R(1) and R(6) resonances. The values of \Trot
are
  directly connected to a cooling rate $\Gamma_{c}$ included in the simulations (presented in Fig. 4). The curves represent three different peak-intensities
    of the laser pulses, whereas the hatched area represents a conservative estimated
     range derived from experimental data as those presented in Fig. 3.}\label{fig 5}
\end{figure}

The ratios between the simulated dissociation rates for REMPD
processes via the R(1) and R(6) resonances are presented in Fig.
\ref{fig 5} as a function of the effective rotational temperature
obtained by including a certain cooling rate $\Gamma_{c}$ in the
calculations.
The dissociation rates have been extracted 10 s after initiation
of the photo-dissociation process where the various populations
and the dissociation rate are simulated to be in steady state
\cite{LangArtikkel}. The chosen dissociation ratio provides a
simple experimental way of distinguishing between low ($\lesssim$
1 K) and high ($\gtrsim$ 100 K) rotational temperatures \Trot as
the ratio changes drastically for temperatures below $\sim$ 50 K.

The ratio of dissociation rates is fortunately nearly insensitive
to intensity variations. Even a factor of four in the actual peak
intensity does not significantly change the ratio. This indicates
that a smaller error in calibration of the power meter used in the
experiments or smaller uncertainties in the laser beam positioning
should not affect the measured ratio. A conservative estimated
interval of this ratio that is in agreement with a series of
experimental data as those presented in Fig. 3 is indicated in
Fig. 5 as the hatched area. From Fig. 5 it is hence reasonable to
conclude that the rotational temperature of the \MgH ions is
higher than 120 K, and likely as high as room temperature, which
is the same rotational temperature found in storage ring
experiments with very weak coupling between the individual ions
\cite{hechtfischer:2809}. In terms of the rotational cooling rate
$\Gamma_{c}$, we can correspondingly conclude that it must be
smaller than $\sim 0.1 $ $\text s^{-1}$.
Since the permanent dipole moment of \MgH (3.6 Debye) is not
particularly small, and its rotational frequencies are not
extraordinary, we expect that molecular ions in Coulomb crystals
do not generally become rotationally cold. Also since the typical
trapping potentials of Penning traps are typically similar to
those of RF traps, no gain with respect to rotational cooling in
such traps is expected.

When experimenting with polar molecular ions at room temperature,
it hence seems necessary to implement a kind of active rotational
cooling in order to achieve state specific molecular ion targets
\cite{vogelius:173003, vogelius0953, vogelius053412}.
Alternatively, one might choose to cool the trap environment to
cryogenic temperatures \cite{Gerlich,brummer:12700}. For non-polar
molecular ions, the coupling between the internal degrees of
freedom and the modes of the Coulomb crystal as well as the BBR
field is smaller than for polar molecules. Hence, such ions
produced by, e.g., a state-selective REMPI-process would possibly
be able to reside in this state through sympathetic cooling, and
could lead to spatial localized state-specific targets.
%\section{Conclusion}

In conclusion, with \MgH ions as the test case, we have proven
that the rotational motion of smaller polar molecular ions
sympathetically cooled into a Coulomb crystal does not become
significantly cooled, but very likely stays in nearly thermal
equilibrium with the surrounding environment.

%\begin{acknowledgments}
We acknowledge Mikael Poulsen, Jan Th{\o}gersen and Henrik
Stapelfeldt for help using the nanosecond lasersystem. We thank
Ivan S. Vogelius for discussions and for providing Einstein
coefficients used in the simulations. The Carlsberg foundation is
gratefully acknowledged for financial support.

%\bibliography{mybibartRot}

\end{document}